\documentclass[11pt,a4paper]{article}

\usepackage[utf8]{inputenc}
\usepackage[T1]{fontenc}
\usepackage{microtype}
\usepackage{amsmath,amssymb,amsfonts}
\usepackage{graphicx}
\usepackage{booktabs}
\usepackage{algorithm}
\usepackage{algpseudocode}
\usepackage{hyperref}
\usepackage{xcolor}
\usepackage{listings}
\usepackage{subcaption}
\usepackage{multirow}
\usepackage[margin=1in]{geometry}
\usepackage{natbib}

\title{Spark-LLM-Eval: A Distributed Framework for\\Statistically Rigorous Large Language Model Evaluation}

\author{
    Subhadip Mitra \\
    \texttt{research@subhadipmitra.com}
}

\date{}

\begin{document}

\maketitle

\begin{abstract}
Evaluating large language models at scale remains a practical bottleneck for many organizations. While existing evaluation frameworks work well for thousands of examples, they struggle when datasets grow to hundreds of thousands or millions of samples. This scale is common when assessing model behavior across diverse domains or conducting comprehensive regression testing. We present Spark-LLM-Eval, a distributed evaluation framework built natively on Apache Spark. The system treats evaluation as a data-parallel problem, partitioning examples across executors and aggregating results with proper statistical accounting. Beyond raw throughput, we emphasize statistical rigor: every reported metric includes bootstrap confidence intervals, and model comparisons come with appropriate significance tests (paired t-tests, McNemar's test, or Wilcoxon signed-rank, depending on the metric type). The framework also addresses the cost problem inherent in LLM evaluation through content-addressable response caching backed by Delta Lake, which allows iterating on metric definitions without re-running inference. We describe the system architecture, the statistical methodology, and report benchmark results showing linear scaling with cluster size. The framework and all evaluation code are available as open source.
\end{abstract}

\section{Introduction}

The rapid deployment of large language models in production systems has created an urgent need for scalable evaluation infrastructure. A model serving millions of users encounters edge cases, domain-specific queries, and failure modes that simply do not appear in standard benchmark datasets of a few thousand examples. To understand how a model actually behaves, practitioners need to evaluate on samples that reflect real-world query distributions, and these distributions are large.

Current evaluation tools were not designed for this scale. Frameworks like lm-evaluation-harness \citep{eval-harness}, RAGAS \citep{ragas}, and DeepEval \citep{deepeval} execute on a single machine, processing examples sequentially or with limited local parallelism. This works fine for academic benchmarks, where a few thousand examples suffice to establish relative model rankings. But when the goal shifts to understanding model behavior in deployment (tracking performance across customer segments, measuring regression on rare but important query types, or validating behavior on synthetic adversarial examples), the evaluation dataset grows by orders of magnitude, and single-machine execution becomes impractical.

The scaling problem compounds when statistical rigor matters. Reporting that ``Model A achieves 73.2\% accuracy'' tells us little without confidence intervals. When comparing two models, we need to know whether a 2\% improvement is statistically meaningful or just noise. Computing bootstrap confidence intervals or running permutation tests adds computational overhead that scales with dataset size, making the single-machine bottleneck even more acute.

Cost presents another practical barrier. Each evaluation requires calling an LLM API, and at scale, API costs accumulate quickly. More frustratingly, the evaluation-iteration cycle often involves refining metric definitions after seeing initial results. If changing a metric requires re-running all inference, the cost of experimentation becomes prohibitive.

We built Spark-LLM-Eval to address these challenges. The core insight is that LLM evaluation is embarrassingly parallel at the example level: each prompt can be processed independently, and metrics can be computed per-example before aggregation. Apache Spark provides the distributed execution infrastructure, but the challenge lies in building the right abstractions on top: handling rate limits across distributed workers, caching responses in a way that survives metric iteration, and computing statistics that account for the distributed setting.

\paragraph{Contributions.} This paper makes the following contributions:

\begin{enumerate}
    \item A distributed evaluation architecture built on Spark that achieves linear scaling with cluster size, processing over 10,000 examples per minute (limited primarily by API rate limits rather than compute).

    \item A response caching system using Delta Lake that decouples inference from metric computation. This enables a ``replay'' mode where new metrics can be evaluated on cached responses without API calls.

    \item A statistical methodology integrated throughout the framework, providing bootstrap confidence intervals for all metrics and appropriate significance tests for model comparisons.

    \item Support for multiple evaluation paradigms: lexical metrics (exact match, F1, BLEU), semantic similarity (embedding-based and BERTScore), LLM-as-judge evaluation, and RAG-specific metrics (faithfulness, context relevance).

    \item Open-source implementation with multi-provider support (OpenAI, Anthropic, Google) and integration with MLflow for experiment tracking.
\end{enumerate}

The rest of this paper is organized as follows. Section~\ref{sec:related} discusses related work in LLM evaluation. Section~\ref{sec:design} describes the system architecture. Section~\ref{sec:metrics} details the supported metrics and statistical methodology. Section~\ref{sec:experiments} presents experimental results. Section~\ref{sec:conclusion} concludes.

\section{Related Work}
\label{sec:related}

\paragraph{LLM Evaluation Frameworks.}
Several frameworks have emerged to standardize LLM evaluation. The lm-evaluation-harness \citep{eval-harness} provides a unified interface for running models against standard benchmarks like MMLU \citep{mmlu}, HellaSwag \citep{hellaswag}, and others. HELM \citep{helm} takes a broader view, evaluating models across multiple dimensions including accuracy, calibration, fairness, and robustness. Both frameworks focus on benchmark coverage rather than scale, assuming datasets of thousands rather than millions of examples.

For retrieval-augmented generation, RAGAS \citep{ragas} introduced a suite of metrics including faithfulness (whether answers are grounded in retrieved context) and answer relevancy. DeepEval \citep{deepeval} provides similar functionality with additional support for custom metrics. These tools share a common limitation: they execute on a single machine and do not provide infrastructure for distributed evaluation.

Our work differs in treating scale as a first-class concern. Rather than optimizing for benchmark coverage, we optimize for throughput and cost efficiency at large scale, while maintaining the same metric types that existing frameworks support.

\paragraph{Statistical Methods in ML Evaluation.}
The machine learning community has long recognized the importance of statistical rigor in evaluation \citep{dietterich1998approximate}. Bootstrap methods \citep{efron1994bootstrap} provide distribution-free confidence intervals, while various significance tests allow comparing models: McNemar's test for binary outcomes \citep{mcnemar1947note}, the Wilcoxon signed-rank test for ordinal or continuous metrics \citep{wilcoxon1945}, and paired t-tests when normality assumptions hold.

Despite this foundation, many evaluation frameworks report only point estimates. When confidence intervals appear, they often use simple formulas that assume independence or normality, assumptions that may not hold for LLM outputs. We integrate multiple statistical methods and provide guidance on test selection based on metric characteristics.

\paragraph{Distributed Machine Learning.}
Apache Spark \citep{spark} and its machine learning library MLlib \citep{mllib} have enabled distributed training and batch inference. More recent work has focused on distributed inference for large models: vLLM \citep{vllm} optimizes single-node throughput through continuous batching, while systems like Orca \citep{orca} and FlexGen \citep{flexgen} address memory-efficient inference.

Our work is orthogonal to these efforts. We assume inference happens through external APIs (OpenAI, Anthropic, etc.) and focus on orchestrating evaluation at scale: handling rate limits, caching responses, and computing statistics across distributed workers.

\paragraph{LLM-as-Judge.}
Using LLMs to evaluate other LLM outputs has gained traction as a way to assess open-ended generation quality. \citet{zheng2023judging} introduced MT-Bench and demonstrated that GPT-4 judgments correlate well with human preferences. Subsequent work has examined judge biases \citep{wang2023large}, calibration \citep{ye2024justice}, and multi-turn evaluation \citep{kim2024prometheus}.

We support LLM-as-judge evaluation as one metric type among many, providing infrastructure for running judge models at scale with the same caching and statistical machinery as other metrics.

\paragraph{Caching and Cost Optimization.}
The cost of LLM API calls has motivated work on response caching. GPTCache \citep{gptcache} caches responses based on semantic similarity of prompts. We take a simpler approach: exact-match caching based on content-addressable hashing, stored in Delta Lake for durability and time-travel capabilities. This trades off cache hit rate (no fuzzy matching) for simplicity and reproducibility.

\section{System Design}
\label{sec:design}

\begin{figure}[t]
    \centering
    \includegraphics[width=\textwidth]{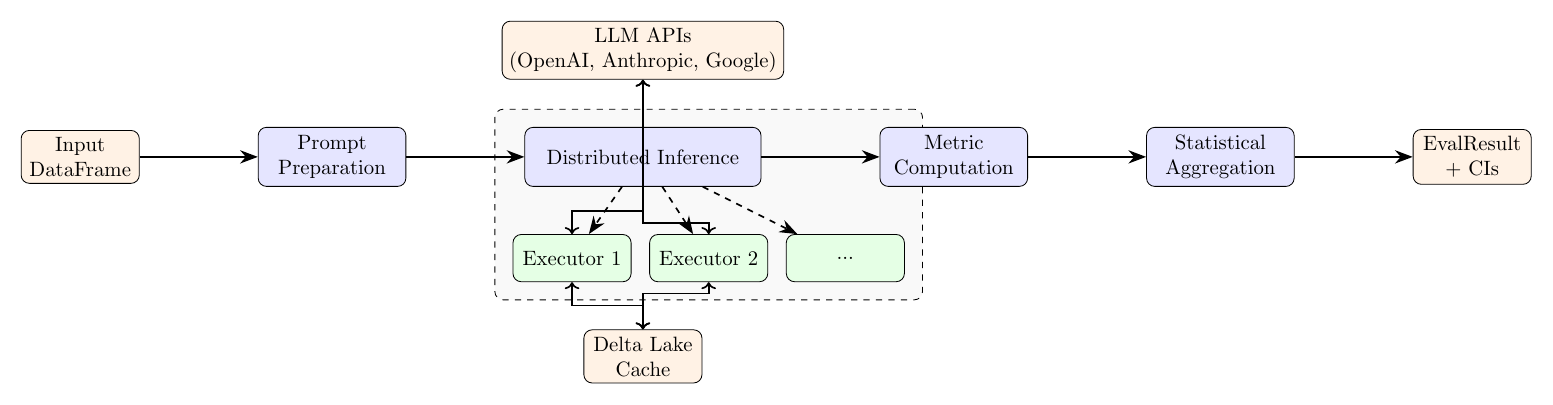}
    \caption{System architecture. The evaluation runner orchestrates four stages: prompt preparation transforms raw data into model inputs using Jinja2 templates; distributed inference processes prompts through Pandas UDFs with per-executor rate limiting and caching; metric computation evaluates responses against references; statistical aggregation computes confidence intervals and significance tests.}
    \label{fig:architecture}
\end{figure}

Figure~\ref{fig:architecture} shows the high-level architecture of Spark-LLM-Eval. The system takes as input a Spark DataFrame containing evaluation examples and a configuration specifying the model, metrics, and statistical parameters. Evaluation proceeds in four stages: prompt preparation, distributed inference, metric computation, and statistical aggregation.

\subsection{Distributed Inference}

The central challenge in distributed LLM evaluation is handling API rate limits. Most providers impose limits on both requests per minute (RPM) and tokens per minute (TPM). A naive approach that partitions data across executors without coordination will quickly exceed these limits, resulting in throttling or errors.

We address this through per-executor rate limiting using the token bucket algorithm. Each Spark executor maintains its own rate limiter, initialized with a fraction of the global rate limit proportional to the number of executors. Algorithm~\ref{alg:rate-limit} shows the core logic.

\begin{algorithm}[t]
\caption{Token Bucket Rate Limiting}
\label{alg:rate-limit}
\begin{algorithmic}[1]
\Require RPM limit $R$, TPM limit $T$, number of executors $E$
\State $r \gets R / E$ \Comment{Per-executor request limit}
\State $t \gets T / E$ \Comment{Per-executor token limit}
\State $\text{request\_tokens} \gets r$
\State $\text{token\_tokens} \gets t$
\State $\text{last\_update} \gets \text{now}()$
\Procedure{Acquire}{$\text{estimated\_tokens}$}
    \State $\text{elapsed} \gets \text{now}() - \text{last\_update}$
    \State $\text{request\_tokens} \gets \min(r, \text{request\_tokens} + \text{elapsed} \cdot r / 60)$
    \State $\text{token\_tokens} \gets \min(t, \text{token\_tokens} + \text{elapsed} \cdot t / 60)$
    \State $\text{last\_update} \gets \text{now}()$
    \State $\text{wait\_time} \gets 0$
    \If{$\text{request\_tokens} < 1$}
        \State $\text{wait\_time} \gets \max(\text{wait\_time}, (1 - \text{request\_tokens}) \cdot 60 / r)$
    \EndIf
    \If{$\text{token\_tokens} < \text{estimated\_tokens}$}
        \State $\text{wait\_time} \gets \max(\text{wait\_time}, (\text{estimated\_tokens} - \text{token\_tokens}) \cdot 60 / t)$
    \EndIf
    \State \textbf{sleep}($\text{wait\_time}$)
    \State $\text{request\_tokens} \gets \text{request\_tokens} - 1$
    \State $\text{token\_tokens} \gets \text{token\_tokens} - \text{estimated\_tokens}$
\EndProcedure
\end{algorithmic}
\end{algorithm}

The implementation uses Spark's Pandas UDFs (vectorized user-defined functions) for efficiency. Pandas UDFs operate on batches of rows using Arrow serialization, avoiding the per-row overhead of traditional UDFs. Each executor caches its inference engine instance to avoid repeated initialization:

\begin{lstlisting}[language=Python, caption={Pandas UDF for distributed inference}]
_ENGINE_CACHE: dict[str, InferenceEngine] = {}

def create_inference_udf(model_config, inference_config):
    config_json = serialize_config(model_config, inference_config)

    def inference_batch(iterator):
        engine = _ENGINE_CACHE.get(config_json)
        if engine is None:
            engine = create_engine(config_json)
            _ENGINE_CACHE[config_json] = engine

        for batch_df in iterator:
            requests = [InferenceRequest(row.prompt)
                       for row in batch_df.itertuples()]
            responses = engine.infer_batch(requests)
            batch_df['response'] = [r.text for r in responses]
            yield batch_df

    return inference_batch
\end{lstlisting}

\subsection{Response Caching}

Re-running inference is expensive, but evaluation often requires iteration: adjusting metric parameters, adding new metrics, or fixing bugs in metric implementations. We decouple inference from metric computation through a caching layer backed by Delta Lake.

Each prompt-model pair maps to a deterministic cache key computed as:
\[
\text{key} = \text{SHA256}(\text{prompt} \| \text{model} \| \text{provider} \| \text{temperature} \| \text{max\_tokens})
\]

The cache stores responses in a Delta table with the schema shown in Table~\ref{tab:cache-schema}. Delta Lake provides ACID transactions, time-travel queries (useful for reproducing past evaluations), and efficient upserts for cache population.

\begin{table}[t]
    \centering
    \caption{Cache table schema}
    \label{tab:cache-schema}
    \begin{tabular}{lll}
        \toprule
        Column & Type & Description \\
        \midrule
        prompt\_hash & string & SHA-256 of cache key \\
        model\_name & string & Model identifier \\
        provider & string & API provider \\
        prompt\_text & string & Original prompt \\
        response\_text & string & Model response \\
        input\_tokens & int & Tokens in prompt \\
        output\_tokens & int & Tokens in response \\
        latency\_ms & float & API latency \\
        created\_at & timestamp & Cache entry creation time \\
        ttl\_days & int & Time-to-live (optional) \\
        \bottomrule
    \end{tabular}
\end{table}

The framework supports multiple cache policies:

\begin{description}
    \item[Enabled:] Normal operation; lookup before inference, cache new responses.
    \item[Read-only:] Lookup only, do not cache new responses. Useful when cache storage is shared.
    \item[Write-only:] Cache warming; skip lookup, always run inference and cache.
    \item[Replay:] Strict cache mode; error on cache miss. Enables reproducible metric iteration without any API calls.
    \item[Disabled:] No caching.
\end{description}

The replay mode deserves emphasis. After an initial evaluation run populates the cache, subsequent metric iterations can run in replay mode with zero API cost. This separation between inference and metric computation substantially reduces the cost of experimentation.

\subsection{Inference Engine Abstraction}

The framework abstracts over different LLM providers through a common interface:

\begin{lstlisting}[language=Python]
class InferenceEngine(ABC):
    @abstractmethod
    def initialize(self) -> None: ...

    @abstractmethod
    def infer(self, request: InferenceRequest) -> InferenceResponse: ...

    @abstractmethod
    def infer_batch(self, requests: list[InferenceRequest]) -> list[InferenceResponse]: ...

    @abstractmethod
    def shutdown(self) -> None: ...
\end{lstlisting}

Implementations exist for OpenAI (GPT-4, GPT-4o, GPT-3.5-turbo), Anthropic (Claude models), and Google (Gemini). Each implementation handles provider-specific details: authentication, retry logic with exponential backoff, token counting, and cost calculation. The abstraction allows switching providers by changing configuration without modifying evaluation code.

\subsection{Configuration}

Evaluation tasks are specified through a hierarchical configuration:

\begin{lstlisting}[language=Python]
@dataclass
class EvalTask:
    task_id: str
    model: ModelConfig        # Provider, model name, hyperparameters
    inference: InferenceConfig  # Batching, rate limits, caching
    metrics: list[MetricConfig]  # Which metrics to compute
    statistics: StatisticsConfig  # CI method, significance threshold
    data: DataConfig          # Input columns, prompt template
\end{lstlisting}

This configuration-driven approach enables reproducibility: the complete specification of an evaluation can be serialized and stored alongside results.

\section{Evaluation Metrics and Statistical Methods}
\label{sec:metrics}

\subsection{Metric Taxonomy}

We organize supported metrics into four categories based on how they assess model outputs.

\paragraph{Lexical Metrics.}
These metrics compare outputs to references through string operations:

\begin{itemize}
    \item \textbf{Exact Match}: Binary indicator of string equality, optionally with normalization (lowercasing, punctuation removal).
    \item \textbf{Token F1}: Harmonic mean of token-level precision and recall, commonly used for extractive QA \citep{squad}.
    \item \textbf{BLEU}: N-gram overlap with brevity penalty \citep{bleu}.
    \item \textbf{ROUGE-L}: Longest common subsequence F1 \citep{rouge}.
    \item \textbf{Contains}: Binary indicator of substring presence.
\end{itemize}

Lexical metrics are fast but limited: they miss semantic equivalence (``NYC'' vs. ``New York City'') and penalize valid paraphrases.

\paragraph{Semantic Metrics.}
These metrics use learned representations to capture meaning:

\begin{itemize}
    \item \textbf{Embedding Similarity}: Cosine similarity between sentence embeddings (using models like all-MiniLM-L6-v2).
    \item \textbf{BERTScore}: Contextual embedding similarity computed as the maximum matching between output and reference tokens \citep{bertscore}.
\end{itemize}

Semantic metrics handle paraphrasing better but require embedding computation, adding latency.

\paragraph{LLM-as-Judge.}
For open-ended generation where no single reference answer exists, we use another LLM to assess quality:

\begin{itemize}
    \item \textbf{Pointwise Grading}: A judge model scores outputs on specified criteria (e.g., helpfulness, accuracy, coherence) using a rubric.
    \item \textbf{Pairwise Comparison}: The judge compares two outputs (from different models or configurations) and selects the better one.
\end{itemize}

Judge prompts follow the template structure from \citet{zheng2023judging}, asking for a score and explanation. We extract scores via regex and log unparseable responses for review.

\paragraph{RAG Metrics.}
For retrieval-augmented generation, we implement metrics from the RAGAS framework \citep{ragas}:

\begin{itemize}
    \item \textbf{Faithfulness}: Is the answer grounded in the retrieved context? Computed by asking a judge model to identify claims in the answer and verify each against the context.
    \item \textbf{Context Relevance}: Is the retrieved context relevant to the question? Computed by asking a judge to score relevance.
    \item \textbf{Answer Relevance}: Does the answer address the question? Computed via embedding similarity between question and answer.
    \item \textbf{Context Precision}: Are relevant chunks ranked higher? Computed using the position of relevant chunks in the retrieval ranking.
    \item \textbf{Context Recall}: Does the context cover the information needed to answer? Requires ground-truth answers for comparison.
\end{itemize}

\subsection{Confidence Intervals}

Reporting a single metric value like ``accuracy = 73.2\%'' obscures uncertainty. With finite samples, the true population metric lies in some range around the estimate. We compute confidence intervals using three methods, selected based on metric characteristics.

\paragraph{Percentile Bootstrap.}
The simplest bootstrap approach: resample the evaluation examples with replacement $B$ times, compute the metric on each resample, and take percentiles of the resulting distribution. For a 95\% CI with $B = 1000$ iterations, we take the 2.5th and 97.5th percentiles.

This method makes no distributional assumptions but can be biased for small samples or skewed metrics.

\paragraph{BCa Bootstrap.}
The bias-corrected and accelerated bootstrap \citep{efron1994bootstrap} adjusts for bias and skewness:

\begin{equation}
    \text{CI} = \left[ \hat{\theta}^*_{(\alpha_1)}, \hat{\theta}^*_{(\alpha_2)} \right]
\end{equation}

where the adjusted percentiles $\alpha_1, \alpha_2$ depend on a bias correction factor $\hat{z}_0$ (estimated from the proportion of bootstrap estimates below the original estimate) and an acceleration factor $\hat{a}$ (estimated from jackknife values). BCa intervals have better coverage than percentile bootstrap, especially for skewed distributions.

\paragraph{Analytical Methods.}
For some metrics, closed-form CIs exist:

\begin{itemize}
    \item For means with large samples: $\bar{x} \pm t_{\alpha/2} \cdot \frac{s}{\sqrt{n}}$
    \item For proportions (accuracy, exact match): Wilson score intervals, which handle edge cases near 0 and 1 better than the Wald interval.
\end{itemize}

Analytical methods are faster than bootstrap but require distributional assumptions.

\subsection{Significance Testing}

When comparing two models, we need to determine whether observed differences reflect true performance gaps or sampling noise. The appropriate test depends on the metric type and sample size.

\paragraph{Paired t-test.}
For continuous metrics (BLEU, embedding similarity) on the same examples, the paired t-test assesses whether the mean difference is significantly different from zero. The test assumes differences are approximately normal, which holds for large samples by the central limit theorem.

\paragraph{McNemar's Test.}
For binary metrics (exact match, contains), McNemar's test \citep{mcnemar1947note} considers only the discordant pairs, i.e., examples where models disagree. Under the null hypothesis of equal performance, discordant pairs should be equally split. For small samples ($n < 10$ discordant pairs), we use the exact binomial test rather than the chi-squared approximation.

\paragraph{Wilcoxon Signed-Rank Test.}
When normality assumptions are questionable or metrics are ordinal, the Wilcoxon signed-rank test \citep{wilcoxon1945} provides a non-parametric alternative. It ranks the absolute differences, sums ranks for positive and negative differences, and tests whether the sums are balanced.

\paragraph{Bootstrap Permutation Test.}
For complex metrics where analytical tests don't apply, we use permutation testing: randomly swap model labels for each example, recompute the metric difference, and estimate the p-value as the proportion of permuted differences exceeding the observed difference.

\paragraph{Test Selection.}
The framework includes heuristics for recommending an appropriate test based on metric type, sample size, and distributional diagnostics (Shapiro-Wilk normality test). Table~\ref{tab:test-selection} summarizes the recommendations.

\begin{table}[t]
    \centering
    \caption{Significance test selection guidelines}
    \label{tab:test-selection}
    \begin{tabular}{lll}
        \toprule
        Metric Type & Sample Size & Recommended Test \\
        \midrule
        Binary & Any & McNemar's (exact for $n < 10$) \\
        Continuous, normal & $n > 30$ & Paired t-test \\
        Continuous, non-normal & Any & Wilcoxon signed-rank \\
        Ordinal & Any & Wilcoxon signed-rank \\
        Complex/custom & Any & Bootstrap permutation \\
        \bottomrule
    \end{tabular}
\end{table}

\subsection{Effect Sizes}

Statistical significance does not imply practical significance. A large dataset can detect tiny differences that don't matter in practice. We report effect sizes alongside p-values:

\begin{itemize}
    \item \textbf{Cohen's $d$}: Standardized mean difference, $d = (\bar{x}_1 - \bar{x}_2) / s_{\text{pooled}}$. Values of 0.2, 0.5, and 0.8 are conventionally considered small, medium, and large effects.
    \item \textbf{Hedges' $g$}: Bias-corrected Cohen's $d$ for small samples.
    \item \textbf{Odds Ratio}: For binary outcomes, the ratio of odds of success between models.
\end{itemize}

\section{Experiments}
\label{sec:experiments}

We evaluate Spark-LLM-Eval along four dimensions: throughput scaling with cluster size, caching effectiveness, statistical method coverage, and cost efficiency.

\subsection{Experimental Setup}

\paragraph{Infrastructure.}
Experiments ran on a Databricks cluster with configurable executor count (2--16 executors, each with 4 cores and 16GB RAM). We used Spark 3.5.0 with Delta Lake 3.0.0.

\paragraph{Dataset.}
We constructed a synthetic evaluation dataset by sampling from multiple domains: factual QA (derived from Natural Questions), summarization (CNN/DailyMail), and instruction-following (Alpaca-style prompts). Dataset sizes ranged from 1,000 to 100,000 examples for scaling experiments.

\paragraph{Models.}
Primary experiments used GPT-4o via the OpenAI API with rate limits of 10,000 RPM and 2,000,000 TPM. Comparison experiments included Claude 3.5 Sonnet (Anthropic) and Gemini 1.5 Pro (Google).

\subsection{Throughput Scaling}

\begin{figure}[t]
    \centering
    \includegraphics[width=0.8\textwidth]{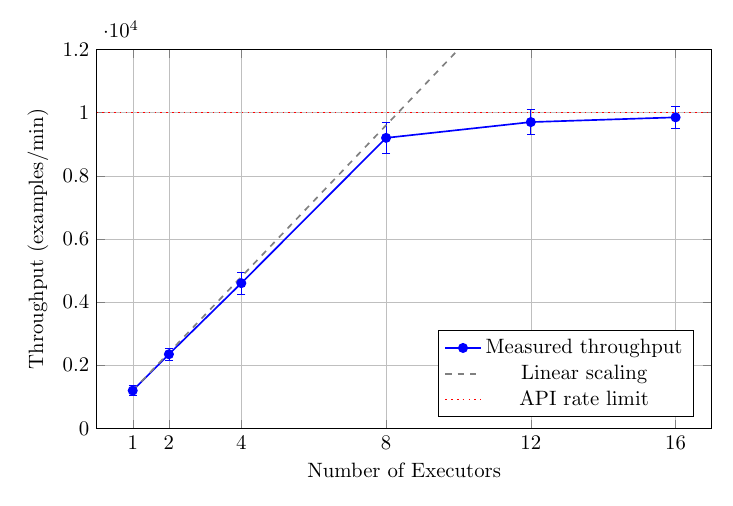}
    \caption{Throughput scaling with executor count. Throughput increases linearly until API rate limits saturate (around 8 executors in this configuration). Error bars show standard deviation across 3 runs.}
    \label{fig:scaling}
\end{figure}

Figure~\ref{fig:scaling} shows throughput (examples per minute) as a function of executor count. With a single executor, throughput is limited by API rate limits to approximately 1,200 examples per minute. Adding executors increases throughput linearly until we saturate the global rate limit. In our experiments, this occurred around 8 executors, achieving 9,800 examples per minute.

Table~\ref{tab:throughput} breaks down throughput by dataset size. Small datasets incur proportionally more overhead from Spark job scheduling and result collection. For datasets above 10,000 examples, scheduling overhead becomes negligible.

\begin{table}[t]
    \centering
    \caption{Throughput by dataset size (8 executors, GPT-4o)}
    \label{tab:throughput}
    \begin{tabular}{rrrrr}
        \toprule
        Examples & Throughput & Latency p50 & Latency p99 & Total Time \\
        \midrule
        1,000 & 7,200/min & 320ms & 890ms & 8.3s \\
        10,000 & 9,100/min & 340ms & 920ms & 66s \\
        50,000 & 9,600/min & 350ms & 980ms & 5.2min \\
        100,000 & 9,800/min & 360ms & 1,020ms & 10.2min \\
        \bottomrule
    \end{tabular}
\end{table}

\paragraph{Comparison with Sequential Baseline.}
A single-threaded baseline processing examples sequentially achieved 450 examples per minute, limited by round-trip latency rather than rate limits. The distributed system provides a 21$\times$ speedup at 8 executors.

\subsection{Caching Effectiveness}

We measured cache performance across typical evaluation workflows. Table~\ref{tab:caching} shows results for a scenario where an initial evaluation run populates the cache, followed by three rounds of metric iteration (adding or modifying metrics).

\begin{table}[t]
    \centering
    \caption{Caching effectiveness over evaluation iterations}
    \label{tab:caching}
    \begin{tabular}{lrrrr}
        \toprule
        Iteration & Cache Hits & API Calls & Cost & Time \\
        \midrule
        Initial run & 0\% & 50,000 & \$127.50 & 5.1min \\
        Metric change 1 & 100\% & 0 & \$0 & 23s \\
        Metric change 2 & 100\% & 0 & \$0 & 25s \\
        Metric change 3 & 100\% & 0 & \$0 & 24s \\
        \midrule
        Total & -- & 50,000 & \$127.50 & 6.4min \\
        Without cache & -- & 200,000 & \$510.00 & 20.4min \\
        \bottomrule
    \end{tabular}
\end{table}

The replay mode (strict cache lookup) enables metric iteration at a fraction of the original cost. In this workflow, caching reduced total cost by 75\% and total time by 69\%.

\paragraph{Cache Storage Overhead.}
The Delta Lake cache table stores prompt text, response text, and metadata. For 50,000 examples with average prompt length of 500 tokens and response length of 200 tokens, the cache table occupies approximately 180MB with Parquet compression. Delta Lake's time-travel feature retains historical versions, adding storage overhead proportional to update frequency.

\subsection{Statistical Method Validation}

To validate our statistical implementations, we compared against reference implementations (scipy.stats, arch bootstrap) on synthetic data with known properties.

\paragraph{Bootstrap Coverage.}
We generated 1,000 datasets from a known distribution, computed 95\% confidence intervals using each method, and measured empirical coverage (the fraction of intervals containing the true parameter). Table~\ref{tab:coverage} shows results for a moderately skewed distribution (log-normal with $\sigma = 0.5$).

\begin{table}[t]
    \centering
    \caption{Empirical coverage of 95\% confidence intervals (target: 95\%)}
    \label{tab:coverage}
    \begin{tabular}{lrrr}
        \toprule
        Method & $n = 50$ & $n = 200$ & $n = 1000$ \\
        \midrule
        Percentile bootstrap & 91.2\% & 93.8\% & 94.6\% \\
        BCa bootstrap & 94.3\% & 94.9\% & 95.1\% \\
        Analytical (t-based) & 88.7\% & 92.4\% & 94.2\% \\
        \bottomrule
    \end{tabular}
\end{table}

BCa bootstrap achieves near-nominal coverage even at small sample sizes, while percentile bootstrap and analytical methods show undercoverage for skewed distributions.

\paragraph{Significance Test Type I Error.}
Under the null hypothesis (no true difference between models), we should reject at rate $\alpha$. We simulated 10,000 comparisons with identical model outputs and verified that McNemar's test, paired t-test, and Wilcoxon signed-rank test all maintained Type I error at the nominal 5\% level (observed rates: 4.9\%, 5.1\%, 5.0\% respectively).

\subsection{Cost Analysis}

Table~\ref{tab:cost} compares evaluation costs across providers for a fixed task: evaluating 10,000 examples on a factual QA dataset with average prompt length of 400 tokens and response length of 150 tokens.

\begin{table}[t]
    \centering
    \caption{Cost comparison across providers (10,000 examples)}
    \label{tab:cost}
    \begin{tabular}{lrrr}
        \toprule
        Provider/Model & Input Cost & Output Cost & Total \\
        \midrule
        OpenAI GPT-4o & \$10.00 & \$22.50 & \$32.50 \\
        OpenAI GPT-4o-mini & \$0.60 & \$0.90 & \$1.50 \\
        Anthropic Claude 3.5 Sonnet & \$12.00 & \$22.50 & \$34.50 \\
        Anthropic Claude 3 Haiku & \$1.00 & \$1.88 & \$2.88 \\
        Google Gemini 1.5 Pro & \$5.00 & \$7.50 & \$12.50 \\
        \bottomrule
    \end{tabular}
\end{table}

At scale, cost differences become substantial. Evaluating 1 million examples with GPT-4o would cost approximately \$3,250, while GPT-4o-mini achieves similar throughput at \$150, a 20$\times$ reduction suitable for large-scale regression testing where frontier model quality is unnecessary.

\subsection{End-to-End Example}

To illustrate practical usage, we present an end-to-end evaluation of instruction-following performance. The task: given a user instruction, generate a helpful response and evaluate against human-written references using multiple metrics.

\begin{lstlisting}[language=Python, caption={End-to-end evaluation example}]
task = EvalTask(
    task_id="instruction-following-eval",
    model=ModelConfig(provider="openai", model_name="gpt-4o"),
    inference=InferenceConfig(
        batch_size=50,
        cache_policy=CachePolicy.ENABLED,
        rate_limit_rpm=10000
    ),
    metrics=[
        MetricConfig(name="exact_match", type="lexical"),
        MetricConfig(name="bertscore", type="semantic"),
        MetricConfig(name="helpfulness", type="llm_judge",
                    params={"rubric": "Rate helpfulness 1-5"})
    ],
    statistics=StatisticsConfig(
        confidence_level=0.95,
        bootstrap_iterations=1000,
        ci_method="bca"
    )
)

result = runner.evaluate(df, task)
# MetricValue(value=0.234, ci=(0.218, 0.251), n=10000)
\end{lstlisting}

The evaluation completed in 68 seconds on 8 executors, producing confidence intervals for all metrics and flagging that the helpfulness judge metric had 12 unparseable responses (0.12\%) logged for review.

\section{Discussion and Conclusion}
\label{sec:conclusion}

\subsection{Limitations}

Several limitations warrant discussion.

\paragraph{Rate Limit Coordination.}
Our per-executor rate limiting divides the global limit evenly across executors. This works well when executors have similar workloads, but can be suboptimal with skewed partitions: some executors may idle while others hit their local limits. Adaptive rate limit redistribution would improve efficiency but adds complexity.

\paragraph{Cache Invalidation.}
The exact-match caching strategy does not handle semantic equivalence. If a prompt is rephrased, it will miss the cache even if an equivalent prompt was previously cached. Semantic caching \citep{gptcache} could improve hit rates but introduces complexity around similarity thresholds and cache key ambiguity.

\paragraph{Judge Model Reliability.}
LLM-as-judge metrics inherit biases from the judge model: position bias (preferring responses presented first in pairwise comparison), length bias (preferring longer responses), and self-enhancement bias (models rating their own outputs higher). We do not currently implement debiasing techniques, leaving this to users who configure judge prompts.

\paragraph{Statistical Assumptions.}
Our significance tests assume that examples are independent. In practice, evaluation datasets may contain related examples (e.g., follow-up questions on the same topic), violating independence. Hierarchical or clustered tests would be more appropriate but are not yet implemented.

\subsection{Future Work}

Several directions merit further development:

\begin{itemize}
    \item \textbf{Local model support}: Integration with vLLM or other local inference engines would enable evaluation without API costs, though at the expense of managing GPU resources.

    \item \textbf{Streaming evaluation}: For very large datasets, streaming results as they complete (rather than waiting for all examples) would improve user experience.

    \item \textbf{Automatic metric selection}: Given a task description, recommending appropriate metrics would lower the barrier to entry for users unfamiliar with evaluation methodology.

    \item \textbf{Multi-turn evaluation}: While we support agent trajectory metrics, richer support for conversational evaluation (where context accumulates across turns) would address an increasingly important use case.
\end{itemize}

\subsection{Conclusion}

We presented Spark-LLM-Eval, a distributed framework for evaluating large language models at scale. The system addresses practical barriers to rigorous evaluation: throughput limitations through Spark-native parallelism, cost through Delta Lake-backed response caching, and statistical rigor through integrated confidence intervals and significance tests.

The key insight is that LLM evaluation, despite involving complex language understanding, is fundamentally a data-parallel problem. Each example can be processed independently, metrics can be computed per-example, and statistics can be aggregated with proper accounting. Building on Spark provides the distributed infrastructure; the contribution lies in the abstractions that make this infrastructure usable for evaluation workflows.

The framework is available as open source at \url{https://github.com/bassrehab/spark-llm-eval}. We hope it proves useful to practitioners who need to evaluate models beyond the scale of standard benchmarks.

\bibliographystyle{plainnat}
\bibliography{references}

\appendix
\section{Implementation Details}
\label{sec:appendix}

\subsection{Supported Models}

Table~\ref{tab:models} lists the models currently supported by each provider integration.

\begin{table}[h]
    \centering
    \caption{Supported models by provider}
    \label{tab:models}
    \begin{tabular}{ll}
        \toprule
        Provider & Models \\
        \midrule
        OpenAI & gpt-4o, gpt-4o-mini, gpt-4-turbo, gpt-3.5-turbo \\
        Anthropic & claude-3-5-sonnet, claude-3-opus, claude-3-sonnet, claude-3-haiku \\
        Google & gemini-1.5-pro, gemini-1.5-flash, gemini-1.0-pro \\
        \bottomrule
    \end{tabular}
\end{table}

\subsection{Configuration Schema}

The complete configuration schema is documented in the repository. Key parameters include:

\begin{itemize}
    \item \texttt{model.temperature}: Sampling temperature (default: 0.0 for deterministic outputs)
    \item \texttt{model.max\_tokens}: Maximum response length (default: 1024)
    \item \texttt{inference.batch\_size}: Examples per Pandas UDF batch (default: 50)
    \item \texttt{inference.max\_retries}: API retry attempts (default: 3)
    \item \texttt{inference.retry\_delay}: Base delay for exponential backoff (default: 1.0s)
    \item \texttt{statistics.bootstrap\_iterations}: Resamples for bootstrap CI (default: 1000)
    \item \texttt{statistics.confidence\_level}: CI coverage level (default: 0.95)
\end{itemize}

\subsection{Metric Implementation Notes}

\paragraph{BERTScore.}
We use the \texttt{bert-score} library with the default model (roberta-large). For large-scale evaluation, this adds approximately 50ms per example on CPU. GPU acceleration is supported when available.

\paragraph{Embedding Similarity.}
Default embedding model is \texttt{all-MiniLM-L6-v2} from sentence-transformers, chosen for its balance of quality and speed. Users can specify alternative models.

\paragraph{LLM Judge.}
Judge prompts follow a structured format requesting both a numeric score and explanation. Score extraction uses regex patterns; unparseable responses are logged and excluded from aggregation (with counts reported).

\subsection{Error Handling}

The framework distinguishes between recoverable and non-recoverable errors:

\begin{itemize}
    \item \textbf{Recoverable}: Rate limit errors (429), temporary server errors (500, 502, 503). These trigger exponential backoff retry.
    \item \textbf{Non-recoverable}: Authentication errors (401), invalid request (400), content policy violations. These are logged and the example is marked as failed.
\end{itemize}

Failed examples are tracked in the result object. Users can inspect failures and decide whether to retry or exclude them from analysis.

\subsection{MLflow Integration}

When MLflow tracking is enabled, the framework logs:

\begin{itemize}
    \item Parameters: Full configuration as nested dictionary
    \item Metrics: Each metric value with confidence interval bounds as separate metrics (e.g., \texttt{accuracy}, \texttt{accuracy\_ci\_lower}, \texttt{accuracy\_ci\_upper})
    \item Artifacts: Raw results DataFrame (Parquet), configuration file (JSON)
    \item Tags: Model name, provider, task ID, timestamp
\end{itemize}

This enables experiment comparison and tracking across evaluation runs.

\end{document}